\documentstyle[12pt]{article}
\input{epsfig.sty}
\newcommand{\beq}{\begin{eqnarray}}
\newcommand{\eeq}{\end{eqnarray}}
\begin{document}
\title{Heavy quark state production in Pb-Pb collisions
 at $\sqrt{s_{pp}}$=5.02 TeV}
\author{Leonard S. Kisslinger\\
Department of Physics, Carnegie Mellon University, Pittsburgh PA 15213 USA.\\
Debasish Das\\
Saha Institute of Nuclear Physics,1/AF, Bidhan Nagar, Kolkata 700064, INDIA.}
\date{}
\maketitle
\begin{abstract}
   We estimate differential rapidity cross sections for
$J/\Psi$, $\Psi(2S)$, $\Upsilon(1S)$, $\Upsilon(2S)$, and $\Upsilon(3S)$ 
production via Pb-Pb  collisions at proton-proton energy $\equiv \sqrt{s_{pp}}$
=5.02 TeV. For the $\Psi(2S)$ and $\Upsilon(3S)$  states we use the mixed heavy
quark hyrid theory and compare the cross section to the standard model. This is
an extension of
previous work on heavy quark state production via Cu-Cu and Au-Au collisions
at $\sqrt{s_{pp}}$=200 GeV

\end{abstract}
\noindent
PACS Indices:12.38.Aw,13.60.Le,14.40.Lb,14.40Nd
\vspace{1mm}


\section{Introduction}

   In anticipation of measurements of the production of  heavy quark states
via Pb-Pb collisions with proton-proton energy $\equiv \sqrt{s_{pp}}$=5.02 TeV at the
LHC, we estimate the
production of $J/\Psi, \Psi(2S)$, $\Upsilon(1S)$, $\Upsilon(2S)$, $\Upsilon(3S)$
states using the standard model, and $ \Psi(2S), \Upsilon(3S)$ states
 using the mixed hybrid theory. This in an extension of the recent work
on heavy quark state production in A-A (Cu-Cu and Au-Au) collisions at
$\sqrt{s_{pp}}$=200 GeV\cite{klm14}, based on earlier work on heavy quark state 
production in p-p collisions\cite{klm11} which used the color octet 
model\cite{cl96,bc96,fl96} 

        In Section 2 we review the production of $\Psi$ and 
$\Upsilon$ states in Pb-Pb collisions, with small modifications from
Ref\cite{klm14}. In Section 3 we discuss the ratio of $\Psi(2S)$ to
$J/\Psi(1S)$ production and compare the hybrid vs standard theory to recent 
experiments with p-p collisions. In section 4 we discuss the ratio of 
$\Upsilon(3S)$ to $\Upsilon(1S),\Upsilon(2S)$ production and compare the
 hybrid to the standard theory, also reviewed in Ref\cite{kd16}

 Now we briefly review the mixed hybrid theory for charmonium and bottomonium
states.  Using the method of QCD sum rules it was shown\cite{lsk09} that
the $\Psi(2S)$ and $\Upsilon(3S)$ are approximately 50-50 mixtures
of standard quarkonium and hybrid quarkonium states:
\beq
\label{hybrid}
        |\Psi(2s)>&=& -0.7 |c\bar{c}(2S)>+\sqrt{1-0.5}|c\bar{c}g(2S)>
\nonumber \\
        |\Upsilon(3S)>&=& -0.7 |b\bar{b}(3S)>+\sqrt{1-0.5}|b\bar{b}g(3S)>
 \; ,
\eeq
with a 10\% uncertainty in the QCD sum rule estimate of the  mixing 
probabiltiy, while the $J/\Psi,\Upsilon(1S),\Upsilon(2S)$ states are 
essentially standard $q \bar{q}$ states. 

\section{$J/\Psi$ and $\Upsilon(1S)$ production in Pb-Pb collisions
 with $\sqrt{s_{pp}}$ = 5.02 TeV}

 The differential rapidity cross section for the production of a heavy
quark state $\Phi$ with helicity $\lambda=0$ (for unpolarized  
collisions\cite{klm11}) in the color octet model in A-A collisions is given 
by\cite{klm14}

\beq
\label{1}
   \frac{d \sigma_{AA\rightarrow \Phi(\lambda=0)}}{dy} &=& 
   R^E_{AA} N^{AA}_{bin}< \frac{d \sigma_{pp\rightarrow \Phi(\lambda=0)}}{dy}>
\; ,
\eeq
where $R^E_{AA}$ is the product of the nuclear modification factor $R_{AA}$
and $S_{\Phi}$, the dissociation factor after the state $\Phi$ is 
formed\cite{star02}). $N^{AA}_{bin}$ is the number of binary collisions in the AA
collision, and $< \frac{d \sigma_{pp\rightarrow \Phi(\lambda=0)}}{dy}>$ is the 
differential rapidity cross section for $\Phi$ production via nucleon-nucleon 
collisions in the nuclear medium. Experimental studies show that $R^E_{AA}\simeq
 0.5$ both for Cu-Cu\cite{star09,phenix08} and Au-Au\cite{phenix07,star07,
kks06}, and we use $R^E_{AA}= 0.5$ for Pb-Pb collisions. The
number of binary collisions\cite{sbstar07} $N^{PbPb}_{bin}\simeq$  260 for Pb-Pb.
Therefore in Eq(\ref{1}) $R^E_{AA} N^{AA}_{bin}\rightarrow R^E_{PbPb} N^{PbPb}_{bin}
\simeq 130$. The differential rapidity cross section for pp collisions 
in terms of $f_g$\cite{CTEQ6,klm11}, the gluon distribution function is

\beq
\label{2}
     < \frac{d \sigma_{pp\rightarrow \Phi(\lambda=0)}}{dy}> &=& 
     A_\Phi \frac{1}{x(y)} f_g(\bar{x}(y),2m)f_g(a/\bar{x}(y),2m) 
\frac{dx}{dy} \; ,
\eeq  
where $a= 4m^2/s$ and  $A_\Phi=\frac{5 \pi^3 \alpha_s^2}{288 m^3 s}
<O_8^\Phi(^1S_0)>$\cite{klm11} with $m=1.5$  GeV for charmonium, and 5 GeV for 
bottomonium. With $\sqrt{s}$=5.02 TeV, $A_\Phi=1.26 \times 10^{-6}$ nb for 
$\Phi$=$J/\Psi$ and $3.4 \times 10^{-8}$ nb for $\Phi$=$\Upsilon(1S)$.
$a =3.6 \times 10^{-7}$ for charmonium and $4.0 \times 10^{-6}$ for bottomonium.
See Ref\cite{klm11} for a detailed discussion of the theoretical basis
of Eq(\ref{2}) for $pp$ collisions in free space, including references to
earlier publications that showed that $^3P_J$ contributions are much smaller
than the $^1S_0$ scalar contributions.  

 The function $\bar{x}$, the effective parton x in a nucleus (A), is given in  
Refs\cite{vitov06,vitov09}:
\beq
\label{barx}
         \bar{x}(y)&=& (1+\frac{\xi_g^2(A^{1/3}-1)}{Q^2})x(y) \nonumber \\
   x(y) &=& 0.5 \left[\frac{m}{\sqrt{s_{pp}}}(\exp{y}-\exp{(-y)})+
\sqrt{(\frac{m}{\sqrt{s_{pp}}}(\exp{y}-\exp{(-y)}))^2 +4a}\right] \;,
\eeq
with\cite{qiu04} $\xi_g^2=.12 GeV^2$. Therefore for Pb with $A\simeq 208$,
\beq
\label{barx2}
         \bar{x}(y)&=& 1.058 x(x) \; .
\eeq

for $\sqrt{s}$=5.02 TeV, $m/\sqrt{s_{pp}}$ = 0.0003, 0.001 for charmonium, 
bottomonium. The gluon distribution function $f_g$\cite{CTEQ6,klm11} is
\beq
\label{fg} 
    f_g(\bar{x}(y),2m)&=& 1334.21-67056.5 \bar{x}(y)+887962.0 (\bar{x}(y))^2
\; .
\eeq

With $\Psi(2S),\Upsilon(3S)$ enhanced by $\pi^2/4$\cite{klm11} the 
differential rapidity cross sections are shown in the following figures.
The absolute magnitudes are uncertain, and the shapes and relative magnitudes 
are our main prediction.
\clearpage

\begin{figure}[ht]
\begin{center}
\epsfig{file=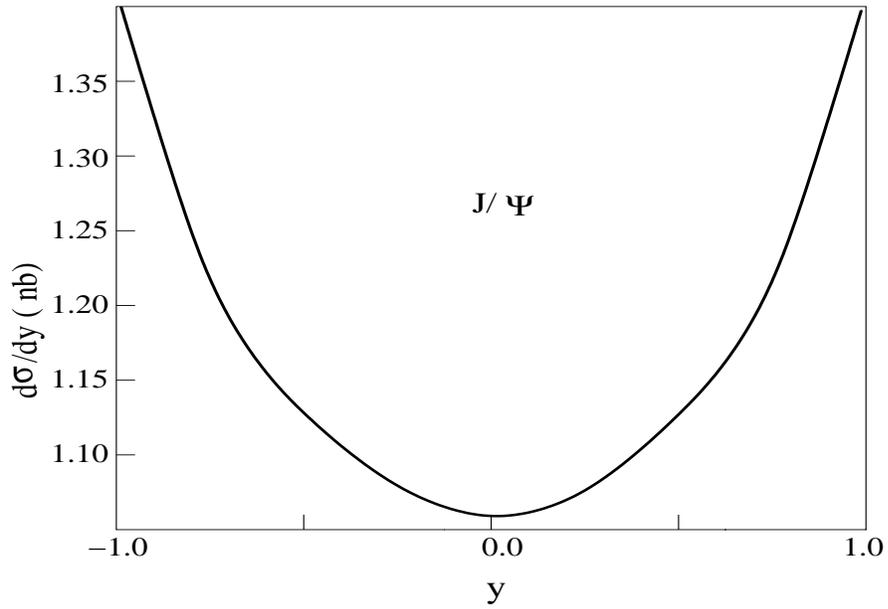,height=8 cm,width=12cm}
\caption{d$\sigma$/dy for 2m=3 GeV, $\sqrt{s_{pp}}$=5.02 TeV Pb-Pb collisions 
producing $J/\Psi$ with $\lambda=0$}
\label{Figure 1}
\end{center}
\end{figure}
\vspace{5.0cm}

\begin{figure}[ht]
\begin{center}
\epsfig{file=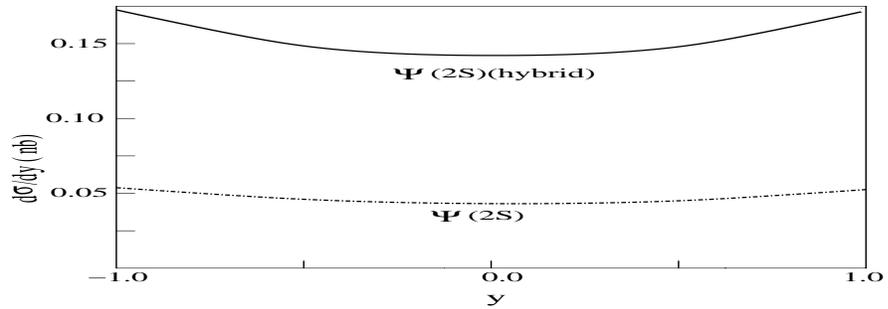,height=4 cm,width=12cm}
\caption{d$\sigma$/dy for 2m=3 GeV, $\sqrt{s_{pp}}$=5.02 TeV Pb-Pb collisions 
producing $\Psi(2S)$, hybrid theory, with $\lambda=0$. The dashed curve is 
for the standard $c\bar{c}$ model.}
\label{Figure 2}
\end{center}
\end{figure}

\clearpage

\begin{figure}[ht]
\begin{center}
\epsfig{file=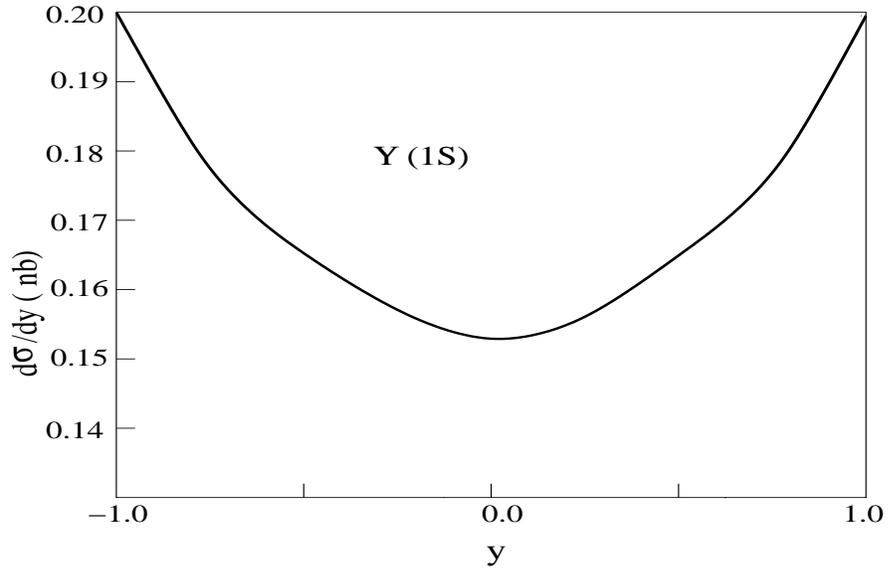,height=7.5 cm,width=12cm}
\caption{d$\sigma$/dy for 2m=10 GeV, $\sqrt{s_{pp}}$=5.02 TeV Pb-Pb collisions 
producing $\Upsilon(1S)$ with $\lambda=0$}
\label{Figure 3}
\end{center}
\end{figure}
\vspace{6.5cm}

\begin{figure}[ht]
\begin{center}
\epsfig{file=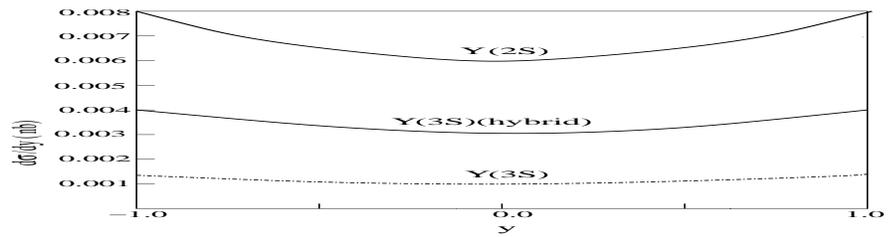,height=3 cm,width=12cm}
\caption{d$\sigma$/dy for 2m=10 GeV, $\sqrt{s_{pp}}$=5.02 TeV Pb-Pb collisions 
producing with $\lambda=0$ $\Upsilon(2S)$ and $\Upsilon(3S)$(hybrid).
For $\Upsilon(3S)$ the dashed curve is for the standard $b\bar{b}$ model.}
\label{Figure 4}
\end{center}
\end{figure}
\clearpage

\newpage

\section{Ratio of $\Psi(2S)$ to $J/\Psi$ cross sections}

In this section we discuss the ratios of the charmonium cross sections
for p-p and Pb-Pb collisions at the LHC. In order to estimate the $\Psi(2S)$
to $J/\Psi$ ratios in Pb-Pb collisions we make use of recent experimental
results on $\Upsilon(mS)$ state production at the LHC.

\subsection{Ratios for p-p collisions} 

  In Ref\cite{klm11} we discussed the $\Upsilon(mS)$ cross section ratios,
showing that the error in the ratios is small as it is given by the wave 
functions for the standard model and the enhancement factor of 
$(1 \pm .1)\times \pi^2/4$ for the mixed hybrid, as discussed in Section 2. 
Now there are accurate 
measurements of the $\Psi(2S)$ to $J/\Psi$ ratio for p-p production at RHIC. 
From the standard (st), hybrid model(hy) one finds for p-p production of 
$\Psi(2S)$ and $J/\Psi$
\beq
\label{ppratio}
    \sigma(\Psi(2S))/\sigma(J/\Psi(1S))|_{st} &\simeq& 0.27 \nonumber \\
    \sigma(\Psi(2S))/\sigma(J/\Psi(1S))|_{hy} &\simeq& 0.67\pm 0.07 \; ,
\eeq
while the PHENIX experimental result for the ratio\cite{phoenix} $\simeq$
0.59. Therefore, as in our earlier work the hybrid model is consistent with
experiment, while the standard model ratio is too small.

\subsection{Ratios for Pb-Pb collisions}

  The recent CMS/LHC result comparing Pb-Pb to p-p Upsilon
production\cite{cms11} found
\beq
\label{CMS2}
     [\frac{\Upsilon(2S) +\Upsilon(3S)}{\Upsilon(1S)}]_{Pb-Pb}/
    [\frac{\Upsilon(2S) +\Upsilon(3S)}{\Upsilon(1S)}]_{p-p} &\simeq& 
0.31^{+.19}_{-.15} \pm .013(syst) \; ,
\eeq
while in our previous work on $p-p$ collisions we found the ratio 
$\sigma(\Upsilon(3S))/\sigma(\Upsilon(1S))|_{p-p}$
of the standard $|b\bar{b}>$ model was $4/\pi^2 \simeq 0.4$ of the hybrid
model. This suggests a suppression factor for $\sigma(b\bar{b}(3S))/
\sigma(b\bar{b}(1S))$, or $\sigma(c\bar{c}(2S)/\sigma(c\bar{c}(1S))$ of 
0.31/.4 as these components travel 
through the QGP; or an additional factor of 0.78 for $\Psi(2S)$ to $J/\Psi$
production for $A-A$ vs $p-p$ collisions. Therefore from Eq(\ref{ppratio}) 
one obtains our estimate using our mixed hybrid theory for this ratio
\beq
  \sigma(\Psi(2S))/\sigma(J/\Psi(1S))|_{Pb-Pb{\rm \; collisions}} &\simeq& 
0.52 \pm 0.05 
\eeq

\section{Ratio of $\Upsilon(2S)$ and $\Upsilon(3S)$
 to $\Upsilon(1S)$ cross sections}

  In our previous work\cite{klm11} we estimated the ratios of $\Upsilon(2S)$ 
and $\Upsilon(3S)$ to $\Upsilon(1S)$ cross sections in comparison with an
experiment published in 1991\cite{fermi91}. Our result for p-p collisions, with
uncertainty due to separating $\Upsilon(2S)$ from $\Upsilon(3S)$, was
\beq
\label{lsk11}
     \Upsilon(3S)/\Upsilon(1S)|_{p-p} &\simeq& 0.14-0.22 \; ,
\eeq
for our mixed hybrid theory, while the standard model would give
$\frac{\Upsilon(3S)}{\Upsilon(1S)} \simeq 0.06$. A recent CMS 
result\cite{cms12}, with a correction factor for acceptance and 
efficiency of the $\Upsilon(3S)$ to the  $\Upsilon(1S)$ state, which was 
estimated to be approximately 0.29\cite{cms11}, was found to be\cite{klm14}
\beq
\label{cms12}
     \Upsilon(3S)/\Upsilon(1S)|_{p-p} &\simeq& 0.12 \; ,
\eeq
with the mixed hybrid theory in agreement within errors, while the standard 
model differs by a factor of two.

  The new CMS experiment's main objective\cite{cms12} is to test for 
$\Upsilon$ suppression in PbPb collisions. A recent (unpublished) CMS 
(preliminary) estimate of $\Upsilon$ ratios is\cite{cms16}
\beq
\label{cmsz12}
  && \frac{[\Upsilon(2S)/\Upsilon(1S)]_{PbPb}}
 {[\Upsilon(2S)/\Upsilon(1S)]_{pp}} \simeq 0.2 {\rm \;to\;} 0.4 \nonumber \\
  &&\frac{[\Upsilon(3S)/\Upsilon(1S)]_{PbPb}}
 {[\Upsilon(3S)/\Upsilon(1S)]_{pp}} \simeq 0.0  {\rm \;to\;} 0.26 \; .
\eeq
  The studies of Pb-Pb collisions for bottomonium states, which cannot be
carried out at RHIC but are an important part of LHC programs,
will be carried out in our future research.

\section{Conclusions}

  We have studied the differential rapidity cross sections for
$J/\Psi, \Psi(2S)$ and $\Upsilon(nS)(n=1,2,3)$ production via Pb-Pb 
collisions at the LHC ($\sqrt{s_{pp}}$=5.02 TeV) using $R^E_{AA}$, the product 
of the nuclear modification factor $R_{AA}$ and the dissociation factor 
$S_{\Phi}$, $N^{AA}_{bin}$ the binary collision number, and the gluon distribution
functions from previous publications. This should give some guidance for 
future LHC experiments, although at the present time the $\Upsilon(nS)$ 
states cannot be resolved.

  The ratio of the production of $\sigma(\Psi(2S))$, which in our mixed
hybrid theory is 50\% $c\bar{c}(2S)$ and 50\% $c\bar{c}g(2S)$ with a 
$10\%$ uncertainty, to
$J/\Psi(1S)$, which is the standard $c\bar{c}(1S)$, will be an important
test of the production of the quark-gluon plasma. Using the hybrid model
and suppression factors from previous theoretical estimates and experiments
on $\Upsilon(mS)$ state production at the LHC, we estimate that the ratio 
of $\Psi(2S)$ to $J/\Psi(1S)$ production at the LHC via Pb-Pb collisions will 
be about $0.52\pm 0.05$.

\vspace{1cm}
\Large{{\bf Acknowledgements}}

\vspace{1cm}
\normalsize 
Author L.S.K. acknowledges support from the P25 group at Los 
Alamos National laboratory. Author D.D. acknowledges the facilities of Saha 
Institute of Nuclear Physics, Kolkata, India.

\end{document}